\documentclass[aps,prd,twocolumn,showpacs,floatfix,preprintnumbers,nofootinbib]{revtex4-1}
\usepackage{graphicx}  
\usepackage{dcolumn}   
\usepackage{bm}        
\usepackage{amssymb}   
\usepackage{amsmath}   
\usepackage{hyperref}  
\usepackage{color}     

\begin{document}

\title{Dilepton constraints in the Inert Doublet Model from Run~1 of the LHC}

\author{Genevi\`eve B\'elanger$^{1}$, B\'eranger Dumont$^{2,3}$, Andreas Goudelis$^{4}$, Bj\"orn Herrmann$^{1}$, Sabine Kraml$^{3}$, Dipan Sengupta$^{3}$}

\affiliation{$^{1}$LAPTh, Universit\'e Savoie Mont Blanc, CNRS, B.P.\ 110, F-74941 Annecy-le-Vieux, France\\
$^{2}$Center for Theoretical Physics of the Universe, Institute for Basic Science (IBS), Daejeon 305-811, 
Republic of Korea\\
$^{3}$Laboratoire de Physique Subatomique et de Cosmologie, Universit\'e Grenoble-Alpes, 
CNRS/IN2P3, 53 Avenue des Martyrs, F-38026 Grenoble, France\\
$^{4}$Institute of High Energy Physics, Austrian Academy of Sciences, Nikolsdorfergasse 18, 1050 Vienna, Austria}

\begin{abstract}
Searches in final states with two leptons plus missing transverse energy, targeting supersymmetric particles or invisible decays of the Higgs boson, were performed during Run~1 of the LHC.
Recasting the results of these analyses in the context of the Inert Doublet Model (IDM) using {\sc MadAnalysis~5}, we show that  they provide constraints on inert scalars that significantly extend previous limits from LEP. 
Moreover, these LHC constraints allow to test the IDM in the limit of very small Higgs--inert scalar coupling, where the constraints from  direct detection of dark matter and the invisible Higgs width vanish. 
\end{abstract}


\maketitle


The Inert Doublet Model is one of the simplest extensions of the Standard Model (SM). 
First introduced as a toy model for electroweak (EW) symmetry breaking studies~\cite{IDMfirst}, 
it was later found to potentially improve naturalness~\cite{IDMnaturalness}, be compatible with 
Coleman-Weinberg EW symmetry breaking (EWSB)~\cite{IDMCW} as well as accommodate dark matter (DM) 
through different potential mechanisms~\cite{IDMnaturalness,IDMarchetype,ScalarMultiplet,IDMnewviable} 
and with interesting signatures~\cite{IDMgammalines,IDMnu1,IDMnu2,IDMpos,IDMVIB}. It can moreover explain 
EW baryogenesis~\cite{IDMbaryogenesis1,IDMbaryogenesis2}, neutrino masses~\cite{IDMnumasses}, 
can be constrained from LEP measurements~\cite{IDMLEPII} and predicts a rich phenomenology at the  
LHC~\cite{IDMnaturalness,IDMLHChinvfirst,IDMdileptons1,IDMtrileptons,IDMmultileptons,IDMhgaga1,IDMhgaga2}.

The discovery of a Higgs-like particle with a mass of about $125$~GeV at the LHC~\cite{ATLASDiscovery,CMSDiscovery}
has already made a tremendous impact on the phenomenology of the IDM, dramatically reducing its available parameter 
space~\cite{IDMposthiggs}. 
Besides its interplay with Higgs phenomenology, the IDM can yield interesting signals at the LHC, involving in particular two~\cite{IDMdileptons1}, three~\cite{IDMtrileptons} or multiple~\cite{IDMmultileptons} 
leptons along with missing transverse energy, $E_{T}^{\rm miss}$. Such signals have so far been studied in the 
literature only as predictions and never as constraints from existing LHC results. 
In this work we show that existing data on $\ell^+\ell^- + E_{T}^{\rm miss}$ searches from the $8$~TeV LHC run 
--- performed by the experimental collaborations with supersymmetry (SUSY) or invisible Higgs decays in mind --- 
begin to provide significant constraints on the IDM parameter space that are highly complementary to those obtained from DM observables.


Let us begin by briefly presenting the IDM and setting some useful notations for the subsequent analysis. 
In the IDM, the SM is extended by the addition of a 
second scalar, $\Phi$, transforming as a doublet under $SU(2)_L$.  
This doublet $\Phi$ is odd under a new discrete $\mathbb{Z}_2$ symmetry, whereas all other fields are even. 
In Feynman gauge we can write the two scalar doublets as 
\begin{equation}
	H = \left( \begin{array}{c} G^+ \\ \frac{1}{\sqrt{2}}\left(v + h + i G^0\right) \end{array} \right),
	\
	\Phi = \left( \begin{array}{c} H^+\\ \frac{1}{\sqrt{2}}\left(H^0 + i A^0\right) \end{array} \right),
\end{equation}
where $v = \sqrt{2}~\langle 0 | H | 0 \rangle \approx 246$ GeV denotes the vacuum expectation value of the neutral component of $H$. 
The $h$ state corresponds to the physical SM-like Higgs boson, whereas $G^0$ and $G^{\pm}$ are the Goldstone bosons. The ``inert'' sector consists of a neutral CP-even scalar $H^0$, a pseudo-scalar $A^0$, and a pair of charged scalars $H^{\pm}$. 
The scalar potential of the model reads
\begin{align}
	V_0 & = \mu_1^2 |H|^2  + \mu_2^2|\Phi|^2 + \lambda_1 |H|^4+ \lambda_2 |\Phi|^4 \\ \nonumber
		& + \lambda_3 |H|^2| \Phi|^2 + \lambda_4 |H^\dagger\Phi|^2 + \frac{\lambda_5}{2} \Bigl[ (H^\dagger\Phi)^2 + \mathrm{h.c.} \Bigr].
\label{Eq:TreePotential}
\end{align}
The masses and interactions of the scalar sector are fixed by the scalar-potential parameters 
$\lambda_{1\ldots5}$ and $\mu_2$, which can be traded for the physically more intuitive set
\begin{equation}
 	\left\{ m_{h}, \ m_{H^0}, \ m_{A^0}, \ m_{H^{\pm}}, \ \lambda_L, \ \lambda_2 \right\},
	\label{eq:masses}
\end{equation} 
where the Higgs and inert scalar masses are given by
\begin{align}
	m_{h}^2 &= \mu_1^2 + 3 \lambda_1 v^2, \\ 	
	m_{H^0}^2 &= \mu_2^2 + \lambda_L v^2, \label{Eq:mH0tree} \\
	m_{A^0}^2 &= \mu_2^2 + \lambda_S v^2, \\
	m_{H^{\pm}}^2 &= \mu_2^2 + \frac{1}{2} \lambda_3 v^2, 
\end{align}
and the couplings $\lambda_{L,S}$ are defined as
\begin{align}
	\lambda_{L,S} &= \frac{1}{2} \left( \lambda_3 + \lambda_4 \pm \lambda_5 \right).
\end{align}
The parameter $\mu_1^2$ is eliminated by using (after EWSB) the scalar potential minimization relation $m_{h}^2 = -2\mu_1^2 = 2 \lambda_1 v^2$.

The $\mathbb{Z}_2$ symmetry forbids mixing among the components of $H$ and $\Phi$ and renders the lightest $\mathbb{Z}_2$-odd particle stable which, if electrically neutral ({\it i.e.}\ $H^0$ or $A^0$), can play the role of a DM candidate. Detailed accounts of the DM phenomenology of the IDM can be found, for example, in \cite{IDMarchetype,ScalarMultiplet,IDMnewviable}. Sticking, for simplicity (but without loss of generality), to the mass hierarchy $m_{H^0} < m_{{A^0}, H^{\pm}}$, the IDM can reproduce the observed \cite{Planck2013} DM abundance according to the freeze-out mechanism in three regimes: the low-mass regime ($m_{H^0} < m_W$), where the relic density is governed by the coupling $\lambda_L$ of two $H^0$  particles to a Higgs boson and the distance of $m_{H^0}$ from $m_h/2$ (the exact difference between $m_{H^0}$ and $m_W$ also plays a role when the former is larger than $\sim 70$ GeV), the intermediate-mass region ($m_W < m_{H^0} \lesssim 115$ GeV), where the relic density depends on $m_{H^0}$ and $\lambda_L$, and the high-mass regime (which will not be of interest for this work) where all parameters of the scalar potential except $\lambda_2$ drastically affect the DM relic abundance. 

The IDM parameter space is subject to numerous constraints that we shall impose throughout our analysis (for a detailed account see, \textit{e.g.}, \cite{IDMposthiggs}). First, there is a minimal set of theoretical requirements (stability of the EW vacuum, perturbativity of all couplings and perturbative unitarity of the scattering matrix) rendering the IDM a consistent and calculable quantum field theory (see also \cite{Khan:2015ipa,Swiezewska:2015paa}). We demand that these three requirements be satisfied up to at least a scale of 10 TeV. We moreover have to ensure that there are no excessive contributions to the oblique parameters $S$, $T$ and $U$, for which we consider the $3\sigma$ ranges from \cite{STUparams}. The analysis presented in \cite{IDMLEPII} showed that (assuming for concreteness $m_{H^0} < m_{A^0}$) the neutralino searches at LEP impose the bound $m_{A^0} \gtrsim 100$ GeV. A recast of the chargino searches at LEP \cite{HchMconstraint} also amounts to a constraint on the charged scalar mass, namely $m_{H^\pm} \gtrsim m_W$. Finally, for $m_{H^0} \leq m_h/2$, we have to make sure that the Higgs boson decay modes predicted by the IDM are in agreement to those observed at the LHC. For SM-like couplings to all SM particles, the invisible branching ratio of a Higgs into two $H^0$ particles is restricted to ${\rm BR}(h\rightarrow {\rm inv.}) < 0.12$ at $95\%$~confidence level (CL)~\cite{StatushCouplingsrun1} (see also \cite{StatusHiggsInv,HiggsAtLast,HiggsSignalStrengths2013,HiggsSignalsStudy}).

Additional constraints arise once the IDM is considered as a DM model. Essentially the entire region where $m_{H^0} < m_h/2$ is ruled out due to the invisible Higgs branching ratio constraint, which imposes $\lambda_L \lesssim 6\times 10^{-3}$, while XENON100 \cite{XENON100225days} already eliminated the entire low-to-intermediate mass regime 
where the IDM can explain DM according to the thermal freeze-out mechanism. The only exception is a narrow region around $m_{H^0} \simeq m_h/2$ which is characterised by $\lambda_L \sim 0$, implying a feeble DM--Higgs (and, hence, DM--nucleon) coupling  \cite{IDMposthiggs}. The more recent LUX results \cite{LUXresults} only render these constraints quantitatively stronger, forcing one to move even deeper into the ``Higgs funnel'' in order to find parameter space points compatible with the relic density constraints from {\it Planck}.

All constraints coming either from invisible Higgs decays or from direct detection experiments however vanish in the limit $\lambda_{L} \rightarrow 0$. Therefore this regime of feeble $\lambda_L$ coupling appears to be extremely challenging experimentally. 
 While it might be argued that, with the exception of the finely-tuned Higgs funnel region, this regime leads to a DM overabundance, it is definitely relevant to seek for constraints on DM models that do not depend in any way upon astrophysical or cosmological assumptions. 
In particular, the ``no-go'' argument for DM overabundance holds only in the context of a ``vanilla'' picture for the thermal history of the universe --- numerous situations can arise in embedings of the IDM which can lead to an eventual dilution of the DM density, along the lines discussed, \textit{e.g.}, in \cite{EntropyInjectionYaguna}. Besides, the IDM needs not be viewed as a DM model, in the sense that it provides an interesting phenomenology in itself. In this spirit, we will see that the dilepton searches at the LHC provide constraints  that are complementary to those obtained from other experimental searches.


The dilepton + $E_{T}^{\rm miss}$ signature in the IDM was first studied in \cite{IDMdileptons1}. 
Four processes provide the main signal contributions to this final state, namely
\begin{align} \label{eq:AH0}
q \bar{q} & \rightarrow Z \rightarrow A^0 H^0 \rightarrow Z^{(*)} H^0 H^0 \rightarrow \ell^+ \ell^- H^0 H^0 , \\ \label{eq:HpHm}
q \bar{q} & \rightarrow Z \rightarrow H^{\pm} H^{\mp} \rightarrow W^{\pm(*)} H^0 W^{\mp(*)} H^0 \\ \nonumber 
          & \rightarrow \nu \ell^+ H^0 \nu \ell^- H^0 , \\ \label{eq:Zh}
q \bar{q} & \rightarrow Z \rightarrow Z h^{(*)} \rightarrow \ell^+ \ell^- H^0 H^0 , \\ \label{eq:fourvertex}
q \bar{q} & \rightarrow Z \rightarrow Z H^0 H^0 \rightarrow \ell^+ \ell^- H^0 H^0 .
\end{align}
Process \eqref{eq:Zh} directly depends on the coupling $\lambda_L$ while all others involve only gauge couplings and thus depend only on the masses of the inert doublet fields.

Searches for two opposite-sign leptons plus $E_{T}^{\rm miss}$ have been performed by both 
the ATLAS and CMS collaborations at Run~1. While no interpretation in the IDM was given by the experiments, there are various Higgs and SUSY analyses that are potentially sensitive to the IDM signatures. 
In the following, we will discuss the results obtained by ATLAS.

On the SUSY side, there is a search for neutralinos, charginos, and sleptons~\cite{ATLASsusy}. 
In this context, the $\ell^+\ell^- + E_{T}^{\rm miss}$ 
signature arises from 
chargino-pair production followed by $\tilde\chi^\pm\to W^{\pm(*)} \tilde\chi^0_1$ or $\tilde\chi^\pm\to \ell^\pm \tilde\nu / \tilde\nu \ell^\pm$ decays, or 
slepton-pair production followed by $\tilde\ell^\pm\to \ell^\pm \tilde\chi^0_1$ decays.
All signal regions (SRs) based on purely leptonic final states require $|m_{\ell\ell}-m_Z| > 10$~GeV, 
{\it i.e.}\ they veto leptons coming from $Z$ decays.  
One of the simplified models considered in this search, $\tilde\chi^+\tilde\chi^- \to W^+(\to \ell^+\nu)\tilde\chi^0_1 W^-(\to \ell^-\nu)\tilde\chi^0_1$, can be directly matched with process~\eqref{eq:HpHm}. Several SRs are designed to optimize sensitivity to different mass splittings $m_{\tilde\chi^\pm}-m_{\tilde\chi^0_1}$.
Interestingly, note that the SUSY equivalent of process~\eqref{eq:AH0}, $\tilde\chi^0_2 \tilde\chi^0_1 \to Z^{(*)}(\to \ell^+\ell^-)\tilde\chi^0_1 \tilde\chi^0_1$, has not been considered in any ATLAS or CMS search. 

On the Higgs side, there is the search for invisible decays of a Higgs boson produced in association with 
a $Z$ boson~\cite{ATLASinvh}. It requires $|m_{\ell\ell}-m_Z| < 15$~GeV, and can be matched to processes~\eqref{eq:Zh} and \eqref{eq:fourvertex}; it may also have sensitivity to process~\eqref{eq:AH0} if $m_{A^0}-m_{H^0} > m_Z$.
Finally, there is another search in the $\ell^+\ell^- + E_{T}^{\rm miss}$ final state, focused on DM~\cite{ATLASdm}. However, it requires $E_{T}^{\rm miss} > 150$~GeV, which completely removes all of our signal.

In order to work out the current LHC constraints on the IDM, we recast these two ATLAS analyses using the 
{\sc MadAnalysis~5}~\cite{Conte:2012fm,Conte:2014zja} framework. The SUSY search~\cite{ATLASsusy}, was already 
available in the Public Analysis Database~\cite{Dumont:2014tja} as the recast code~\cite{ma5susy}. 
The invisible Higgs search~\cite{ATLASinvh} was implemented and validated for this Letter 
and is available at~\cite{ma5higgs}.
The signal generation is done with {\sc MadGraph}~5~\cite{Alwall:2011uj,Alwall:2014hca} 
with model files generated using the {\sc FeynRules} IDM implementation presented in~\cite{IDMposthiggs}; 
the particle widths are calculated with {\sc CalcHEP}~\cite{Pukhov:2004ca,Belyaev:2012qa}. 
The parton-level events are passed through {\sc Pythia}~6.4~\cite{Sjostrand:2006za} for parton 
showering and hadronization before being processed with the `MA5tune' version of {\sc Delphes}~3~\cite{delphes3} 
(see Section~2.2 of \cite{Dumont:2014tja}) for the simulation of detector effects.  
The number of events after cuts are then evaluated with the recast codes~\cite{ma5susy,ma5higgs}. 
For the statistical interpretation, we make use of the module {\tt exclusion\_CLs.py}~\cite{Dumont:2014tja}: 
given the number of signal, observed and expected 
background events, together with the background uncertainty,  {\tt exclusion\_CLs.py} determines 
the most sensitive SR, the exclusion confidence level using the $\mathrm{CL}_s$ prescription, 
and the nominal cross section $\sigma_{95}$ that is excluded at 95\%~CL.\footnote{Note that we do not 
simulate the backgrounds but take the background numbers and uncertainties  directly from the 
experimental publications \cite{ATLASsusy} and \cite{ATLASinvh}.} 

The IDM parameter space, see Eq.~\eqref{eq:masses}, is sampled taking into account the following considerations. 
First, in light of the constraints discussed above which require $\lambda_L$ to be tiny, process \eqref{eq:Zh} is essentially irrelevant for the entire analysis. We can therefore choose $\lambda_L = 0$ without loss of generality. Besides, $\lambda_2$ is irrelevant for all observables at tree-level. 
The mass of the charged inert scalar is important mostly for process \eqref{eq:HpHm}, which comes with the price of an additional EW coupling factor with respect to \eqref{eq:AH0} and turns out to be numerically insignificant unless 
$m_{H^\pm}$ is very light.\footnote{Process \eqref{eq:fourvertex} is also subdominant because the $ZZH^0H^0$ coupling is quadratic in the weak coupling whereas the coupling $ZA^0H^0$ is only linear.} 
We are thus left with $m_{A^0}$ and $m_{H^0}$  to scan over. 
For $m_{H^\pm}$, we choose two representative values: $m_{H^\pm} = 85$~GeV, which is the lower allowed limit by LEP, 
and $m_{H^\pm} = 150$~GeV, which is significantly higher but still safely within the bounds imposed by the $T$ parameter, which limits the mass splitting between the inert scalar states (see also the analysis in \cite{IDMposthiggs}).

\begin{figure}
\includegraphics[width=9cm]{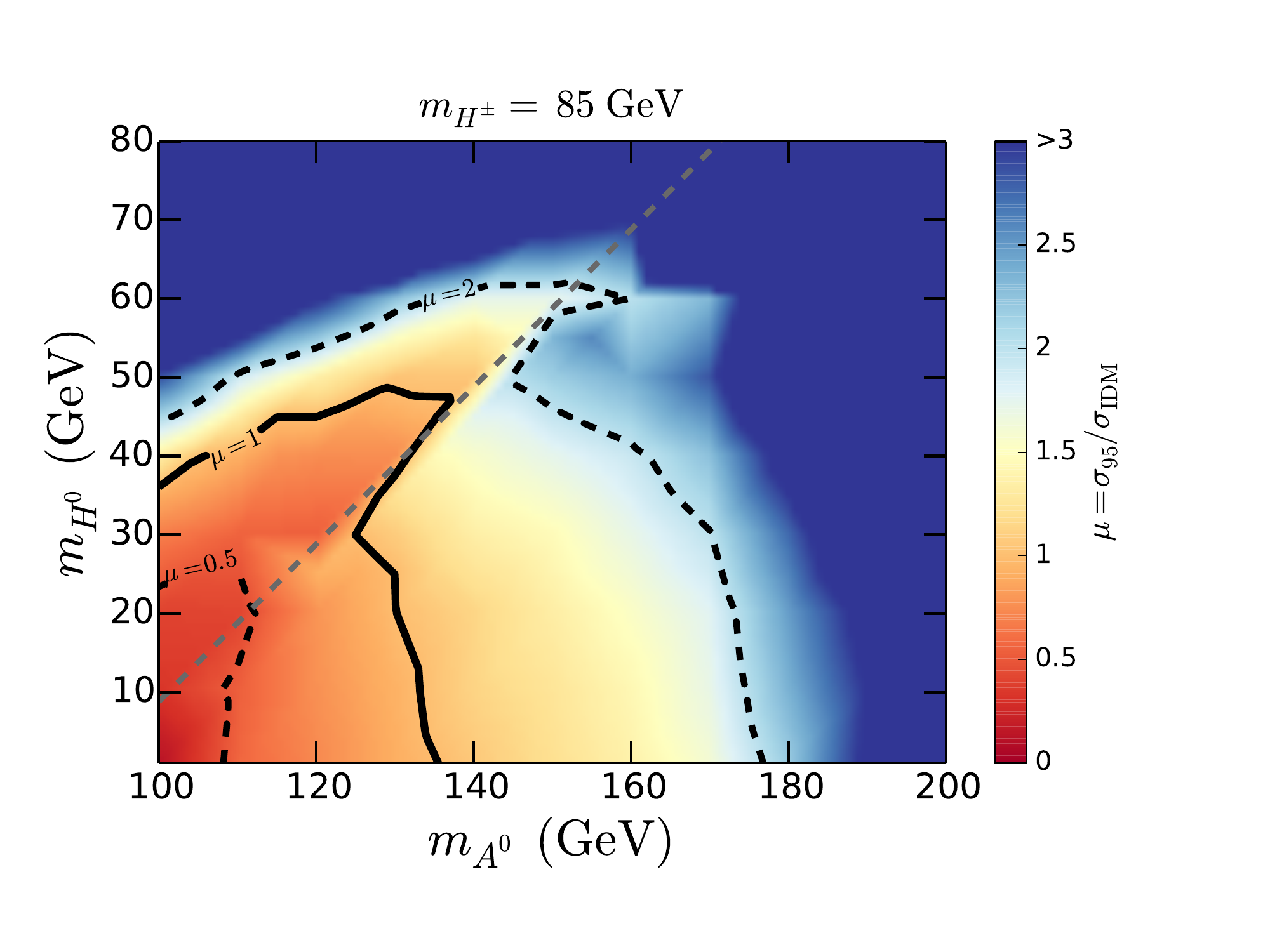}\vspace*{-10mm}
\includegraphics[width=9cm]{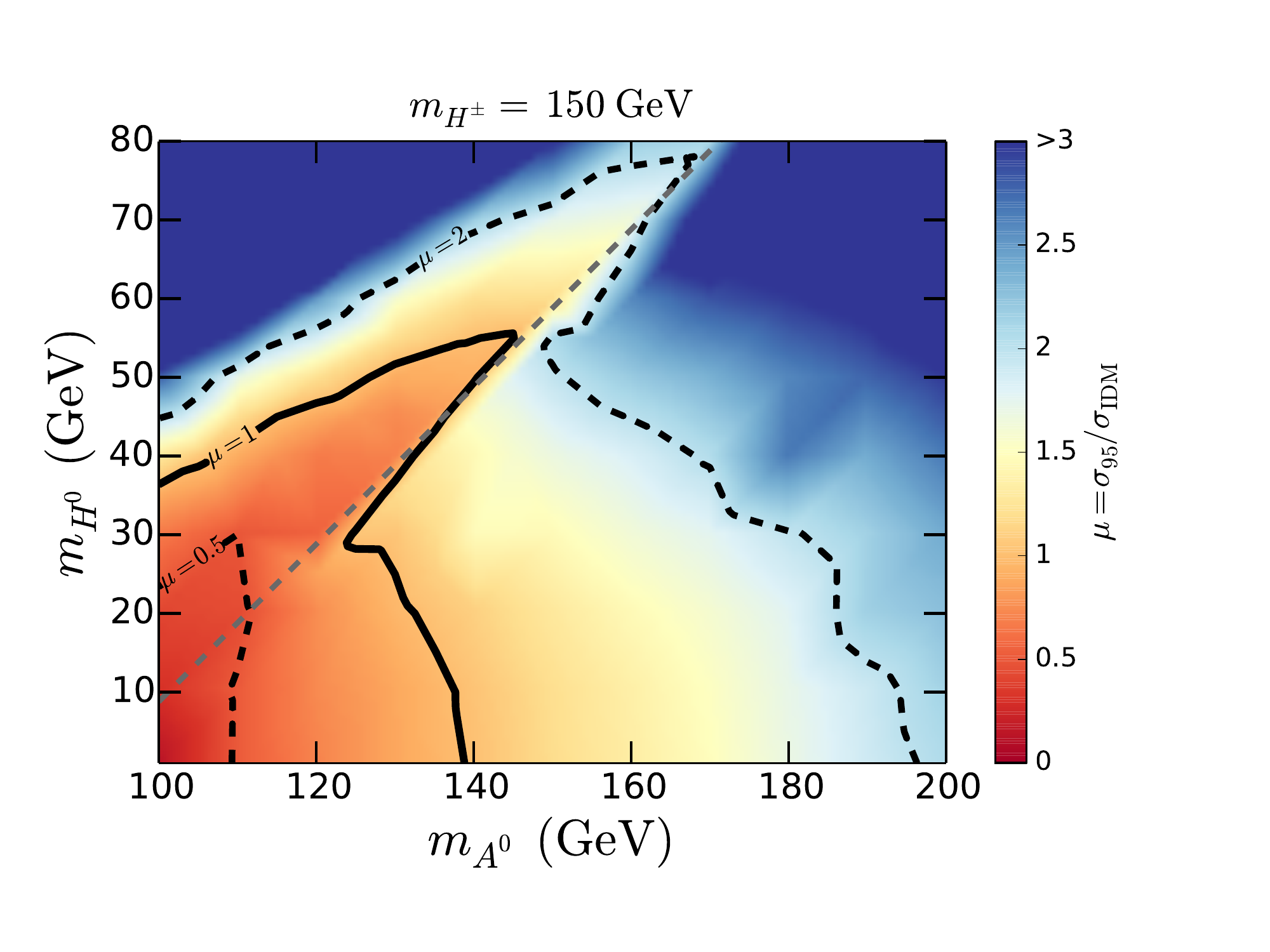}\vspace*{-8mm}
\caption{The ratio $\mu \equiv \sigma_{95}/\sigma_{\rm IDM}$ in the $(m_{A^0}, m_{H^0})$ plane for two representative values of the charged inert scalar mass, $m_{H^\pm} = 85$ GeV (upper panel) and $m_{H^\pm} = 150$ GeV (lower panel). The solid black lines are the 95\%~CL exclusion contours, $\mu = 1$. The dashed black lines are given for illustration and correspond to the $\mu = 0.5$ and $\mu = 2$ contours. The grey dashed lines indicate $m_{A^0}-m_{H^0}=m_Z$. }
\label{fig:exclusionplots}
\end{figure}

The main results of our analysis are presented in Figure~\ref{fig:exclusionplots}, where we show 
$\mu \equiv \sigma_{95}/\sigma_{\rm IDM}$
in the form of temperature plots in the $(m_{A^0}, m_{H^0})$ plane for the two chosen values of $m_{H^\pm}$. 
Here, $\sigma_{\rm IDM}$ is the cross section predicted by the model while $\sigma_{95}$ is the cross section 
excluded at 95\%~CL.   With this definition, regions where $\mu \le 1$ are excluded at 95\%~CL.

As can be seen, the Run~1 ATLAS dilepton searches exclude, at 95$\%$~CL, inert scalar masses up to about $35$~GeV for pseudoscalar masses around $100$~GeV, with the limits becoming stronger for larger $m_{A^0}$, reaching $\approx 45$ (55)~GeV for $m_{A^0} \approx 140$ (145)~GeV and $m_{H^\pm} = 85$ (150)~GeV. 
For massless $H^0$, $A^0$ masses up to about 135--140~GeV are excluded  
(note that $m_{H^0}$ and $m_{A^0}$ are generally interchangeable here).
Several interesting features merit some discussion. 

First, we observe that the constraints are slightly stronger for heavier charged scalars.  
This is in part due to the small contribution from process~\eqref{eq:HpHm} and from $q \bar{q} \to W^\pm \to AH^{\pm} \to Z^{(*)}H W^{\pm(*)}H$ where one of the leptons is missed: although the cross section is much larger for  $m_{H^\pm} = 85$~GeV as compared to $m_{H^\pm} = 150$~GeV, the resulting leptons are much softer and almost never pass the signal requirements. 
A more significant difference between the $m_{H^\pm} = 85$~GeV and 150~GeV cases arises from the fact that at large $m_{A^0}$ the signal from process~\eqref{eq:AH0} is suppressed by the decay $A^0 \to W^{\pm(*)} H^\mp$ followed by $H^\mp \to W^\mp H^0$, that competes with $A^0 \to Z^{(*)} H^0$. While the former decay mode also leads to dileptons, these leptons are, as above, much softer and almost never pass the signal requirements.
For these reasons, the $m_{H^\pm}$ dependence is well captured by Figure~\ref{fig:exclusionplots}; increasing $m_{H^\pm}$ further to, \textit{e.g.}, $300$~GeV has very little effect.\footnote{In the strict IDM framework, $m_{H^\pm}=300$~GeV is excluded by EW precision tests for the $(m_{H^0}, m_{A^0})$ region depicted in Figure~\ref{fig:exclusionplots}, but can be allowed in extensions of the model like in \cite{Belanger:2014bga}.}

Second, and more importantly, the limits on $m_{H^0}$ become stronger for larger $A^0$ masses (above the diagonal line of $m_{A^0}=m_{H^0}+m_Z$). This is also a kinematical effect, because the leptons originating from the $A^0\to Z^{(*)}H^0$ decay become harder when there is more phase space available and thus pass the signal selection cuts more easily. Similarly, for small mass splittings between $H^0$ and $A^0$, the produced dileptons are softer for smaller $m_{A^0}$. 

\begin{figure}
\includegraphics[width=9cm]{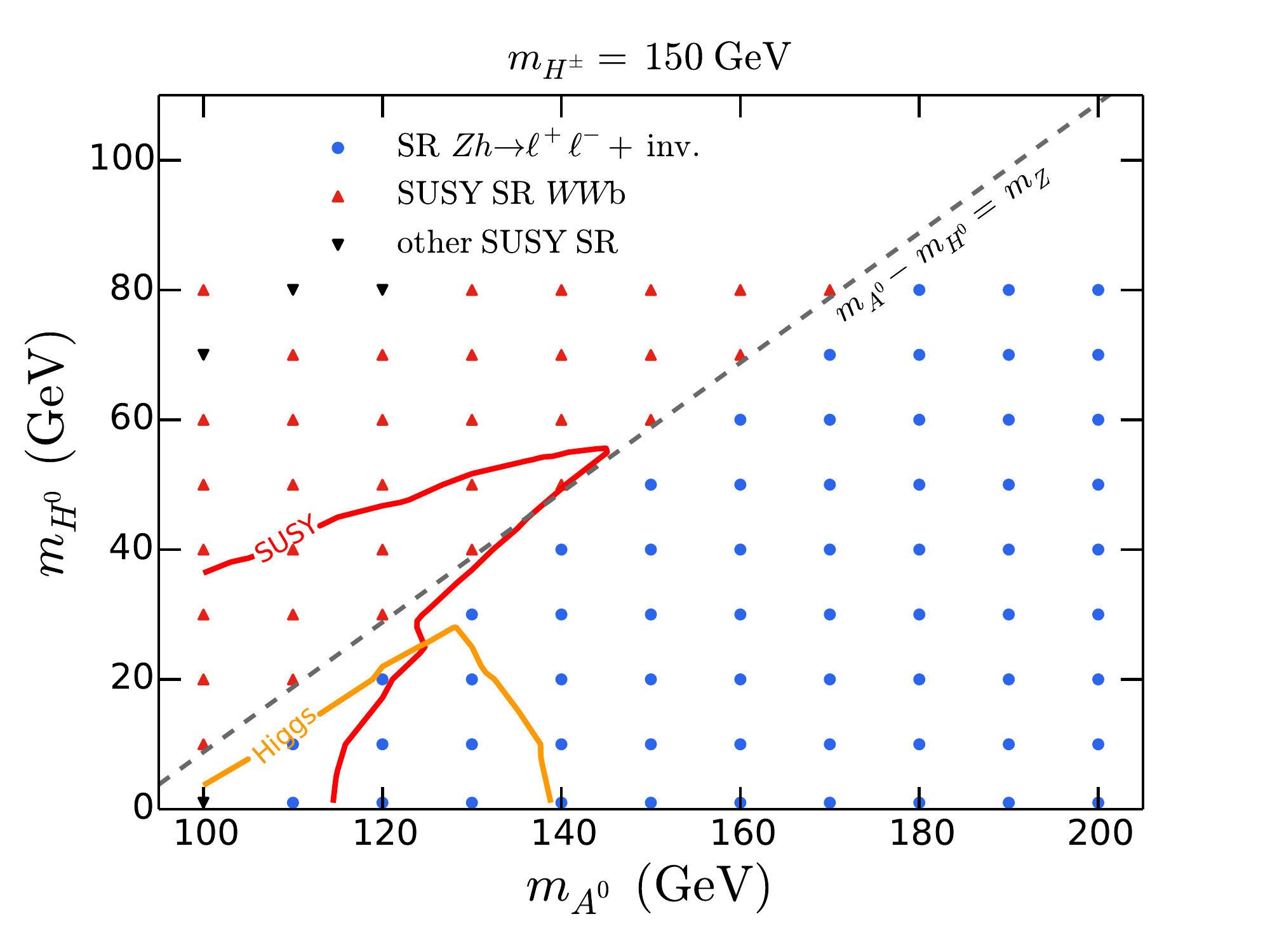}
\caption{Most sensitive SR in the $(m_{A^0}, m_{H^0})$ plane. Also shown as solid lines are the individual 95\%~CL exclusion curves obtained from the SUSY (in red) and Higgs (in orange) analyses considered in this study.}
\label{fig:signalregions}
\end{figure}

Third, once the mass splitting  ($m_{A^0}-m_{H^0}$) is large enough for the $Z$ to be on-shell, the $Z$ veto in the SUSY analysis eliminates most of the signal. In this region it is the $Zh\to \ell^+\ell^- + E_T^{\rm miss}$ search that gives the stronger limit. 
This is illustrated in Figure~\ref{fig:signalregions}, which shows the most sensitive SRs in  the $(m_{A^0}, m_{H^0})$ plane. The concrete example used is for $m_{H^\pm}=150$~GeV but the results basically do not depend on the 
mass of the charged scalar. As expected, for $m_{A^0}-m_{H^0}\le m_Z$ the SUSY dilepton search is more sensitive, while for $m_{A^0}-m_{H^0}> m_Z$ exclusion is driven by the $Zh,h\to {\rm inv.}$ search. 
Also shown are the individual 95\%~CL exclusion curves obtained from the SUSY and Higgs analyses considered in this study. It is worth noting that there is a small overlap region at low $H^0$ mass, $m_{H^0}\lesssim 25$~GeV,  where the SUSY search also gives a limit despite the $Z$ boson from the $A^0$ decay being on-shell.  This is due to a sizeable tail of the $m_{\ell\ell}$ distribution extending below $m_{\ell\ell}=m_Z-10$~GeV that allows to pick up some signal events in the SUSY SR $WWb$ (this tail was already noticed in \cite{IDMdileptons1}). 

The evolution of these constraints in the next run of the LHC is a non-trivial question.  A na\"ive rescaling of signal and background numbers (see, \textit{e.g.}, \cite{Bharucha:2013epa}), assuming that the acceptance$\times$efficiency values remain the same, gives that at 13~TeV and an integrated luminosity of 100~fb$^{-1}$, the 95\%~CL reach should extend up to $\mu\approx1.2$ (1.6) above (below) the dashed grey line in Figure~\ref{fig:exclusionplots}. This improves to $\mu\approx2.1$ (2.7) with  300~fb$^{-1}$, thus covering a major part of 
the $(m_{A^0}, m_{H^0})$ plane shown in Figure~\ref{fig:exclusionplots}. 
The exact reach will depend on many unknown factors, including the performance of the LHC accelerator and of the detectors at Run 2, improvements in the identification and reconstruction algorithms, a better estimation of the SM background, and modified selection cuts. 
The important point we want to make is that even with analyses that are not directly targeted at the IDM, the coverage of the parameter space is quite good. On the other hand, 
the LHC sensitivity could certainly be improved with an analysis optimized for $pp\rightarrow A^0H^0\rightarrow Z^{(*)} H^0H^0$ (or alternatively $pp\rightarrow \tilde\chi_2^0\tilde\chi_1^0\rightarrow Z^{(*)}\tilde\chi_1^0\tilde\chi_1^0$).
In particular, angular separation variables as discussed in \cite{IDMdileptons1} should prove useful to enhance the sensitivity.  We therefore encourage the experimental collaborations to perform a dedicated search for inert scalars at 13~TeV.


In conclusion, the results from Run~1 of the LHC 
provide relevant constraints on the IDM model that significantly extend previous limits from LEP. 
They are complementary to the cosmological and astrophysical constraints, 
as they allow to test the model in the limit $\lambda_L\to 0$, where the constraints from the direct detection of DM and the invisible Higgs width vanish. It is also worth noting that these limits are independent of DM considerations, that may be different for extensions of the model like in \cite{Belanger:2014bga,Banik:2014cfa,Ko:2014uka,Hindmarsh:2014zba,Bonilla:2014xba}. 
The 95\%~CL limits we derived from the dilepton $+\; E_T^{\rm miss}$ Higgs and SUSY analyses exclude inert scalar masses of up to about 55~GeV in the best cases.
It should be possible to push these constraints beyond $m_{h}/2\approx 62.5$~GeV at Run~2 of the LHC, thus starting  to probe the Higgs funnel region of the IDM, a regime that is very hard to test through direct DM searches. 
We stress that these conclusions were reached based on existing analyses that were not optimised at all for the IDM signal. A dedicated experimental analysis for the IDM at 13~TeV would be highly desirable.

\bigskip

{\bf Acknowledgements:} BD and SK thank Maria Krawczyk and Dorota Sokolowska for useful discussions. 
This work was supported in part by the ANR project {\sc DMAstroLHC}, ANR-12-BS05-0006, the 
``Investissements d'avenir, Labex ENIGMASS'', and by the IBS under Project Code IBS-R018-D1.
AG\ is supported by the New Frontiers program of the Austrian Academy of Sciences.

\bibliography{IDMdileptonsRefs}

\end{document}